\documentclass[preprint,aps,prd,nofootinbib,preprintnumbers,amsmath,amssymb]{revtex4}
\usepackage{graphicx}
\usepackage{dcolumn}
\usepackage{bm}
\usepackage{natbib}
\usepackage{color}
\usepackage{slashed}

\newcommand{\nc}{\newcommand}
\newcommand\fft[2]{\frac{#1}{#2}}
\newcommand\ft[2]{{\textstyle\frac{#1}{#2}}}
\newcommand\nn{{\nonumber}}
\newcommand{\beq}{\begin{equation}}
\newcommand{\eq}{\end{equation}}

\nc{\bea}{\begin{eqnarray}} \nc{\ea}{\end{eqnarray}} \nc{\be}{\begin{equation}} \nc{\ee}{\end{equation}} \nc{\barr}{\begin{array}}
\nc{\earr}{\end{array}}
\newcommand{\btop}[2]{\genfrac{[}{]}{0pt}{}{\,#1\,}{\,#2\,}}

\begin{document}

\title{Comments on $a$-maximization from gauged supergravity}

\author{Phillip Szepietowski}
\email{pgs8b@virginia.edu}
\affiliation{Department of Physics, University of Virginia,
Box 400714, Charlottesville, VA 22904, USA}

\begin{abstract}

In this paper we study the holographic dual to $a$-maximization in five-dimensional $\mathcal N = 2$ gauged supergravity. In particular, we apply the procedure described by Tachikawa in \cite{Tachikawa:2005tq} to specific examples consisting of holographic duals to gauge theories arising as the IR limit of $N$ M5-branes wrapping a Riemann surface. A key element of this analysis is a consistent truncation of seven-dimensional $\mathcal N = 4$ $SO(5)$ gauged supergravity reduced on a Riemann surface. We demonstrate the consistency of this truncation and match to a sector of five-dimensional matter-coupled $\mathcal N = 2$ gauged supergravity. 
We determine the $U(1)_R$ symmetry and central charge of these theories and find agreement with the literature. The final results provide a nontrivial illustration of the holographic interpretation of $a$-maximization.


\end{abstract}

\maketitle

\section{Introduction}

$a$-maximization \cite{Intriligator:2003jj} provides an important tool in analyzing supersymmetric conformal field theories (SCFTs.) At an IR superconformal fixed point this simple procedure allows one to identify the unique $U(1)_R$ symmetry whose current sits in the stress-energy tensor supermultiplet. Knowledge of the correct $U(1)_R$ provides useful nonperturbative knowledge of the IR SCFT; for example, it can be used to identify the conformal dimensions of chiral primary operators.

AdS/CFT duality \cite{Maldacena:1997re,Gubser:1998bc,Witten:1998qj} has taught us that many conformal field theories admit a dual large $N$ description in terms of string/M-theory in backgrounds which are products of anti de Sitter space and some internal manifold. Isometries of the internal manifold are mapped to global symmetries in the CFT. In particular for duals to SCFTs the $U(1)_R$ is often given by an appropriate linear combination of $U(1)$ isometries of the internal geometry. An analogue of $a$-maximization in string theory, termed $Z$-minimization, was made precise for Sasaki-Einstein reductions of IIB supergravity in \cite{Martelli:2005tp,Martelli:2006yb}\footnote{And furthermore, proven to be equivalent to $a$-maximization in \cite{Butti:2005vn,Eager:2010yu}.}. This method identifies the appropriate isometry -- termed the Reeb vector of the Sasaki-Einstein manifold -- by minimizing the volume of the internal manifold with respect to all possible choices for this vector.

Supersymmetric $AdS_5$ critical points also exist in M-theory, in which case the internal manifold is six-dimensional and not Sasaki-Einstein. Furthermore, $\mathcal N = 2$ gauged supergravities with $AdS_5$ critical points often arise as consistent truncations of Kaluza-Klein reductions of maximal ten and eleven dimensional supergravity theories \cite{Cvetic:1999xp,Liu:2000gk,Buchel:2006gb,Gauntlett:2006ai,Gauntlett:2007sm,Cassani:2010uw,Liu:2010sa,Gauntlett:2010vu,Skenderis:2010vz,Cassani:2010na,Bena:2010pr,Halmagyi:2011yd}. Such truncations provide the holographic dual description of certain sub-sectors of four dimensional $\mathcal N = 1$ SCFTs. In the correspondence, global symmetries in the gauge theory often arise as gauge symmetries in the supergravity. As such, there should be a dual description of $a$-maximization formulated purely within five-dimensional gauged supergravity. This would be useful for studying more generic dualities. For example one should be able to interpret an $\mathcal N = 2$ gauged supergravity at a supersymmetric $AdS_5$ critical point as being dual to some large $N$ SCFT, even if a specific string theory embedding is lacking. In the cases where a string/M-theory background does exist such a technique would also prove useful for identifying the appropriate isometry of the internal space dual to the $U(1)_R$.

The appropriate description of $a$-maximization in $\mathcal N = 2$ gauged supergravity was determined in \cite{Tachikawa:2005tq}. In this note we wish to illustrate the usefulness of this prescription by working out the $U(1)_R$ for a particular set of $AdS_5$ solutions in eleven-dimensional supergravity. These solutions provide the holographic duals to the low energy limit of M5-branes wrapped on Riemann surfaces. The first examples of these backgrounds were studied in \cite{Maldacena:2000mw,Gaiotto:2009gz} corresponding to duals of $\mathcal N = 2$ SCFTs. Certain $\mathcal N = 1$ generalizations of these constructions have also been studied in \cite{Benini:2009mz}. Recently, expanding on the construction in \cite{Maldacena:2000mw}, an infinite family of solutions dual to $\mathcal N = 1$ SCFTs, which contains some of the solutions described in \cite{Benini:2009mz}, was found in \cite{Bah:2011vv,Bah:2012dg}. For these backgrounds the $U(1)_R$ was determined by utilizing explicit knowledge of the SCFT of the M5-brane theory and matching to the canonical structure of $AdS_5$ solutions in M-theory \cite{Gauntlett:2004zh}, but a dual description of $a$-maximization purely based in supergravity was lacking. Here we provide that description.

A key element in implementing the dual $a$-maximization procedure will be determining a consistent truncation of seven-dimensional $\mathcal N = 4$ $SO(5)$ gauged supergravity (which arises in the dimensional reduction of eleven-dimensional supergravity on a four-sphere.) We do this by reducing the seven-dimensional theory on a Riemann surface to yield a subset of a five-dimensional $\mathcal N = 2$ gauged supergravity. This is similar to the consistent truncations corresponding to duals of other types of wrapped M5-branes discussed in \cite{Gauntlett:2006ai,Gauntlett:2007sm,Donos:2010ax}. The reduction which we present yields an infinite family of supersymmetric consistent truncations -- one corresponding to each unique solution found in \cite{Bah:2012dg}. Additionally, we find many non-supersymmetric truncations. These consistent truncations may prove useful in other holographic studies of the wrapped M5-brane theories.

This paper is organized as follows. In section \ref{sec:review} we review the technique of $a$-maximization and its dual description in five dimensional supergravity. In section \ref{sec:contrunc} we present a consistent truncation of seven-dimensional $\mathcal N = 4$ $SO(5)$ gauged supergravity to a particular sector of a five-dimensional $\mathcal N = 2$ gauged supergravity and discuss various properties of the reduced theory. In section \ref{sec:amax} we present our results on $a$-maximization in the supergravity dual to the wrapped M5-brane theory. Finally, we end in section \ref{sec:conc} with some remarks on relations to the $a$-theorem and potential future directions.

\section{Supergravity dual of $a$-maximization}\label{sec:review}

When an $\mathcal N = 1$ SCFT admits a large $N$ description in terms of $\mathcal N = 2$ gauged supergravity at an $AdS_5$ critical point, the supergravity analogue of $a$-maximization was described in \cite{Tachikawa:2005tq}. In this section we will give a brief overview of those results, following closely the discussion of \cite{Tachikawa:2005tq}.

In an $\mathcal N = 1$ supersymmetric gauge theory one can define a global $U(1)$ symmetry which acts on the fermionic coordinates $\theta_\alpha$ as
\begin{equation}\label{eq:Rtrans}
\theta_\alpha \rightarrow e^{i \phi^I Q_I} \theta_\alpha  = e^{i \phi^I \hat{P}_I} \theta_\alpha,
\end{equation}
where $I$ labels each $U(1)$ symmetry, $\phi^I$ are a set of $U(1)$ transformation parameters, $Q_I$ are the generators of each $U(1)$ and $\hat P_I$ is the charge\footnote{The unusual notation $\hat P_I$ for the $U(1)$ charges of $\theta_\alpha$ is chosen to make comparison with \cite{Tachikawa:2005tq} obvious and in anticipation of its analogue in the supergravity description.} of $\theta_\alpha$ under each $U(1).$ Parameterizing the angles $\phi^I$ by a single angle $\phi,$ so that $\phi^I = s^I\phi$ with some constant coefficients $s^I,$ one normalizes the charge of $\theta_\alpha$ so that $s^I\hat P_I = 1.$ For a choice of such coefficients the $U(1)$ generated by the linear combination $s^I Q_I$ is termed an $R$-symmetry of the theory. The $R$-symmetry defined above is not unique. For any flavor $U(1)$ symmetry given by $t^IQ_I,$ where $t^I$ are some other set of constants with the property that $t^I \hat P_I = 0$ (so that it does not rotate the $\theta_\alpha,$) the linear combination $(s^I + t^I)Q_I$ also acts as an $R$-symmetry.
%
%


However, if the theory is also conformal (an SCFT) there exists a preferred $U(1)_R$ symmetry whose current sits in the stress-energy tensor supermultiplet. {\it A priori} this superconformal $U(1)_R$ symmetry could be given by any linear combination of a naive $R$-symmetry and any $U(1)$ flavor symmetries in the theory. The principle of $a$-maximization proposed in \cite{Intriligator:2003jj} asserts that the correct linear combination of $U(1)$ charges which corresponds to this preferred $U(1)_R$ is the particular linear combination which maximizes the trial $a$-function defined by
\begin{equation}\label{eq:afunction}
a(s) = \frac{3}{32}(3\,\textrm{tr} R(s)^3 - \textrm{tr} R(s)),
\end{equation}
where $R(s) = s^IQ_I$, over all possible choices for the coefficients $s^I$ consistent with (\ref{eq:Rtrans}).
At the maximum, $R(s)$ becomes the superconformal $U(1)_R$ symmetry and the trial $a$-function reduces to the $a$-type central charge of the SCFT \cite{Anselmi:1997am,Anselmi:1997ys}.


\subsection{Five-dimensional $\mathcal N = 2$ supergravity}

Before discussing the prescription of \cite{Tachikawa:2005tq} for implementing this procedure in the supergravity description we will first discuss the relevant details of five-dimensional gauged supergravity.

In general, five-dimensional $\mathcal N=2$ supergravity may be coupled to $n_v$ vector, $n_t$ tensor and $n_h$ hypermultiplets \cite{Ceresole:2000jd}.  We will not consider tensor multiplets as they will not appear in our discussion of a-maximization.  As is well known, the bosonic field content of this theory consists of the metric $g_{\mu\nu}$, $n_v+1$ vectors $A_\mu^I$ (with $I=0,\ldots,n_v$), $n_v$ vector multiplet scalars $\phi^x$ (with $x=1,\ldots,n_v$) parameterizing a so-called ``very special manifold" and $4n_h$ hyperscalars $q^X$ (with $X=1,\ldots,4n_h$) parameterizing a quaternionic manifold.

The bosonic $\mathcal N=2$ Lagrangian is
\begin{eqnarray}\label{eq:N2lag}
\mathcal L &=& R - \fft12 g_{xy}D_\mu\phi^x D^\mu\phi^y - \fft12 g_{XY}D_\mu q^X D^\mu q^Y - V  \nn \\
&& - \fft14 G_{IJ}F^I_{\mu\nu}F^{J\,\mu\nu} + \fft1{24}c_{IJK}\epsilon^{\mu\nu\rho\lambda\sigma}F^I_{\mu\nu}F^J_{\rho\lambda}A^K_\sigma,
\end{eqnarray}
where $V$ is the scalar potential, $g_{xy}$ and $g_{XY}$ are metrics on the very special and quaternionic manifolds, respectively, $c_{IJK}$ are a set of constant coefficients and $G_{IJ}$ is determined by the very special geometry.

The fermionic supersymmetry transformations for the gravitino $\psi_{\mu\,i},$ gauginos $\lambda^x_i$ and
hyperinos $\zeta^A$ are
\begin{eqnarray}\label{eq:N2vars}
\delta\psi_{\mu\,i}&=&\bigl[D_\mu +\ft{i}{24} X_I(\gamma_\mu{}^{\nu\rho} - 4\delta_\mu^\nu\gamma^\rho)F^I_{\nu\rho}\bigr]\epsilon_i + \ft{i}6 \gamma_\mu X^I (P_I)_i{}^j\epsilon_j\,, \nn \\
\delta\lambda^x_i &=&\bigl(-\ft{i}2\gamma^\mu D_\mu\phi^x - \ft14g^{xy}\partial_y X^I\gamma^{\mu\nu}F_{I\,\mu\nu}\bigr)\epsilon_i - g^{xy}\partial_y X^I(P_I)_i{}^j\epsilon_j\,, \nn \\
\delta\zeta^A &=& f^{i\,A}_X\bigl(-\ft{i}2\gamma^\mu D_\mu q^X + \ft12 X^I K^X_I\bigr)\epsilon_i\,,
\label{eq:susy}
\end{eqnarray}
where $i,j=1,2$ are $SU(2)$ indices, $A=1,\ldots,2n_h$ is an $Sp(2n_h)$ index, $f^{i\,A}_X$ are vielbein components on the quaternionic manifold and $\epsilon_i$ is a symplectic-Majorana spinor.

The covariant derivatives are
\begin{eqnarray}
D_\mu\phi^x &=& \partial_\mu\phi^x + A^I_\mu K_I^x(\phi^x)
\end{eqnarray}
for the vector multiplet scalars and
\begin{eqnarray}
D_\mu q^X &=& \partial_\mu q^X + A^I_\mu K_I^X(q^X)
\end{eqnarray}
for the hypermultiplet scalars, where we have fixed the gauge coupling to be unity. The Killing vectors $K_I^x(\phi^x)$ and $K_I^X(q^X)$ correspond to the gauging of the isometries of the very special manifold and quaternionic manifold, respectively.

The vector multiplet scalars are given in terms of $n_v + 1$ constrained scalars
$X^I = X^I(\phi^x)$ subject to the very special geometry constraint
\begin{equation}
\fft16c_{IJK}X^IX^JX^K = 1.
\end{equation}

Additionally, the scalar metric for the vector multiplet scalars is determined by
\begin{eqnarray}\label{eq:vecscmetric}
G_{IJ} &=& X_IX_J - c_{IJK}X^K\,, \nn\\
X_I &=& \fft12 c_{IJK}X^JX^K\,, \nn\\
g_{xy} &=& \partial_x X^I \partial_y X^JG_{IJ}\,.
\end{eqnarray}

The Killing prepotentials $(P_I)_{i}^{\ j}=P^r_I(i\sigma^r)_i^{\ j},$ with $\sigma^r$ the usual Pauli matrices, are determined through differential relations by the Killing vectors and depend only on the hyperscalars. It is the constrained scalars $X^I$ and the prepotentials $P_I^r$ which will show up prominently in the discussion of $a$-maximization (the significance of the scalars of very special geometry also make a related appearance in \cite{Barnes:2005bw}.)

For the present discussion we will restrict to theories in which $(P_I)_i{}^j = P^3_I (\sigma_3)_i{}^j$ and define $P_I \equiv P^3_I.$ 
With this assumption the superpotential defined by
\begin{equation}
P = X^I P_I
\end{equation}
acts as a true superpotential in the sense discussed in \cite{Halmagyi:2011yd} and yields the scalar potential via the relation
\begin{equation}
V = 2 (\mathcal G^{-1})^{ij}\partial_i P\partial_j P - \fft43 P^2,
\end{equation}
where $i,j$ label all of the scalars in the theory and $(\mathcal G^{-1})^{ij}$ is the inverse scalar metric. For the specific example we study in this paper the assumption that $(P_I)_i{}^j = P^3_I (\sigma_3)_i{}^j$ will hold true. But this assumption is imposed only for simplicity, 
the case when it is not true is also discussed in \cite{Tachikawa:2005tq} to which we refer the reader for details.

\subsection{The dual description of $a$-maximization}

The dual a-maximization prescription in \cite{Tachikawa:2005tq} is to identify the charges $\hat P_I$ of the $\theta_\alpha$ under the superconformal $U(1)_R$ with the prepotentials $P_I$ evaluated at the $AdS_5$ critical point and the $s^I$ coefficients of the trial R-charge with the constrained vector scalars $X^I$, normalized such that $s^I P_I = 1$ at the critical point, so that one makes the identification
\begin{equation}
s^I = \frac{X^I}{P}.
\end{equation}

The supergravity analogue of the trial $a$-function is then identified with the inverse cube of the superpotential as follows
\begin{equation}
a(s) \propto \frac{1}{6}c_{IJK}s^Is^Js^K = \frac{\ft16c_{IJK}X^IX^JX^K}{P^{3}} = \frac{1}{P^{3}}.
\end{equation}
The $c_{IJK}$ coefficients play the role of the anomaly coefficients in the gauge theory \cite{Witten:1998qj} which follows from their appearance in the Chern-Simons term in (\ref{eq:N2lag}). Note that this $a$-function only incorporates the $\textrm{tr} R(s)^3$ term in (\ref{eq:afunction}) and not the linear term $\textrm{tr} R(s)$. This is related to the fact that the $\textrm{tr} R(s)$ term derives from a mixed $U(1)$-gravitational-gravitational anomaly in the gauge theory \cite{Anselmi:1997am,Anselmi:1997ys}. In $AdS/CFT$ this comes from higher derivative terms of the form $A^I \wedge \textrm{tr} (R \wedge R)$, where $R$ is the Riemann curvature two-form, which is suppressed at large $N$ \cite{Anselmi:1998zb,Aharony:1999rz,Hanaki:2006pj,Cremonini:2008tw,Liu:2010gz}.

In the absence of nontrivial hypermultiplets, by which we mean that the Killing vectors in the hyperscalar covariant derivatives vanish at the critial point $\tilde{K}_I^X = 0$ (where a tilde is used to imply a quantity has been evaluated at the critical point,) the maximization of the $a$-function is then equivalent to {\it minimizing} the superpotential $P$ with respect to the scalars in the vector multiplets.

In the case with nontrivial hypermultiplets at the critical point, i.e. when $\tilde{K}_I^X \neq 0,$ a slight generalization of this procedure is necessary. In this case the superpotential encodes the generalized $a$-maximization procedure with lagrange multipliers advocated in \cite{Kutasov:2003ux}, which incorporates anomalous global $U(1)$ symmetries. In supergravity, this procedure is necessary when the hypermultiplets are gauged in such a way that in the $AdS_5$ vacuum one or more of the gauge fields are Higgsed -- as seen by a nonvanishing Killing vector. This will be the case in the example studied here. To implement this one separates the superpotential as follows
\begin{equation}
P = X^I P_{(0)\,I} + P_a m^a_I X^I,
\end{equation}
where $P_{(0)\,I}$ and $m^a_I$ are constants and all of the the hypermultiplet dependence is encoded in $P_a.$ The $P_a$ then act as the dual of the lagrange multipliers. One is then instructed to minimize the superpotential with respect to both the vector multiplet scalars $X^I$ as well as the hypermultiplet scalars $q^X$, which are encoded in $P_a.$

The conditions for $a$-maximization described in this section are precisely the conditions for the existence of a supersymmetric $AdS_5$ solution in the first order formalism utilizing the superpotential $P$. Therefore, it is apparent that $a$-maximization is implemented implicitly on the gravity side when finding supersymmetric $AdS_5$ vacua, and one can simply identify the superconformal $U(1)_R$ from the supergravity data. Away from the $AdS_5$ critical points the superpotential acts as the generalized a-function with lagrange multipliers proposed in \cite{Kutasov:2003ux}, which is defined along the RG flow to the critical point.

In the following sections we will illustrate this procedure by identifying the superconformal $U(1)_R$ symmetry in holographic duals of field theories arising as the IR limit of M5-branes wrapping Riemann surfaces of arbitrary genus $g.$ Furthermore, using this example we will demonstrate that this technique can be utilized to determine the appropriate $U(1)$ isometry corresponding to the geometrization of the $U(1)_R$ within a reduction from string/M-theory. First, we must determine the specific five-dimensional supergravity theory which provides the holographic description of the wrapped M5-brane field theory.

\section{Consistent truncations and duals to M5-branes on Riemann surfaces}\label{sec:contrunc}

The field theory on a stack of $N$ flat M5-branes in the large $N$ limit is described by M-theory on $AdS_7\times S^4$ \cite{Maldacena:1997re}. The low energy limit of M-theory is eleven-dimensional supergravity which, after performing a Kaluza-Klein (KK) reduction on $S^4,$ admits a truncation to maximal $\mathcal N = 4$ $SO(5)$ gauged supergravity in seven-dimensions \cite{Pilch:1984xy,Pernici:1984xx}.  It has been explicitly demonstrated that this maximally supersymmetric theory exists as a consistent truncation of the KK reduction from eleven-dimensions \cite{Nastase:1999cb,Nastase:1999kf}. We will analyze a further consistent truncation of this maximally supersymmetric theory, meaning that any solution within the truncation can be uplifted to a solution of eleven-dimensional supergravity.

\subsection{A Consistent Truncation to five dimensional gauged supergravity}

There exists a relatively simple $U(1)^2$ truncation of the $\mathcal N = 4$ $SO(5)$ theory in seven dimensions \cite{Liu:1999ai}.\footnote{The interested reader can refer to appendix \ref{app:N4trunc} for details on this truncation and the following dimensional reduction.} The bosonic sector of the truncation keeps two $U(1)$ gauge fields $A^{(1)}$ and $A^{(2)}$, two scalars $\lambda_1$ and $\lambda_2$, and a single three-form potential $S_3.$ 
The truncation follows from a reduction of eleven dimensional supergravity on a squashed $S^4$ with metric
\begin{equation}\label{eq:metricS4}
ds_{11}^2 = (\pi N \ell_p^3)^{2/3}\bigg[\Delta^{1/3} ds_7^2 + \ft14 \Delta^{-2/3}\bigg(e^{4\lambda_1+4\lambda_2}d\mu_0^2+\sum_{i=1}^2 e^{-2\lambda_i}(d\mu_i^2+\mu_i^2(d\phi_i + 4A^{(i)})^2)\bigg)\bigg],
\end{equation}
where $\ell_p$ is the eleven-dimensional Planck length, the $\mu_{0,1,2}$ satisfy $\mu_0^2 + \mu_1^2 + \mu_2^2 = 1$ and parameterize a two-sphere and
%
%
%
\begin{equation}
\Delta = \sum^2_{i=0} Y_i \mu_i^2,
\end{equation}
with
\begin{equation}
Y_0 = e^{-4\lambda_1-4\lambda_2}, \qquad Y_1 =e^{2\lambda_1}, \qquad Y_2 = e^{2\lambda_2}.
\end{equation}
In the above metric $A^{(1)}$ and $A^{(2)}$ gauge two $U(1)$ isometries of the $S^4,$ the scalars $\lambda_1$ and $\lambda_2$ arise as squashing modes and the three-form potential $S_3$ derives from the eleven-dimensional four-form $G_4$ (the explicit form of $G_4$ can be found in \cite{Cvetic:2000ah}.) In the reduction to five-dimensions the existence of two $U(1)$ gauge fields will imply two global $U(1)$ symmetries in the dual theory, one linear combination of these corresponds to the $U(1)_R.$ Furthermore, if the other linear combination is a flavor symmetry, $a$-maximization is required in order to identify the appropriate $U(1)_R$ in the dual theory.

\subsubsection{Dimensional Reduction}

To make connection with the solutions in \cite{Maldacena:2000mw,Gaiotto:2009gz,Benini:2009mz,Bah:2011vv,Bah:2012dg} dual to wrapped M5-branes we would like to reduce this theory on a Riemann surface in such a way that the resulting theory realizes a five-dimensional gauged supergravity theory. The metric reduction is simply
\begin{equation}\label{eq:metricansatz}
ds_7^2 = e^{-\frac{4B}{3}} ds_5^2 + e^{2B} ds_\Sigma^2,
\end{equation}
where $B$ is a scalar which depends only on the coordinates in $ds_5^2$ and  $ds_\Sigma^2$ is the metric on the Riemann surface. We choose the Riemann surface to be closed and of constant curvature $\kappa \pm 1,\, 0.$ Such surfaces are labeled by their genus $g$. For $g=0$ the surface has $\kappa = 1$ and the metric is given by
\begin{equation}
ds_{\mathbb S^2}^2 = 4\frac{(dx_1^2 + dx_2^2)}{(1 + x_1^2 + x_2^2)^2},
\end{equation}
where $(x_1, x_2)$ takes values in $\mathbb R^2,$ corresponding to the two-sphere. For $g=1$ we have the flat torus with $\kappa = 0$ and metric
\begin{equation}
ds_{\mathbb T^2}^2 = 4\pi(dx_1^2 + dx_2^2),
\end{equation}
where $x_1, x_2 \in [0,1].$
Finally, for $g>1$ the Riemann surface has negative curvature $\kappa = -1$ and the metric is given by
\begin{equation}
ds_{\mathbb H^2}^2 = \frac{1}{x_2^2}(dx_1^2 + dx_2^2).
\end{equation}
In this last case $(x_1,x_2)$ parameterize the upper-half plane $\mathbb H^2$ which is subsequently quotiented by an appropriate Fuchsian subgroup $\Gamma \subset PSL(2,\mathbb R),$ attaining a closed Riemann surface of genus $g > 1.$ Note that in our conventions the volume of the Riemann surface is given by $\frac{4\pi(1-g)}{\kappa},$ where for the torus we take $\frac{(1-g)}{\kappa} = 1.$

For the reduction of the seven-dimensional gauge field strength we take (for $i = 1,2$)
\begin{eqnarray}
F^{(i)} &=& F^{(i)} + p_i vol_\Sigma\,,
\end{eqnarray}
where $vol_\Sigma$ is the volume form on the Riemann surface and $p_i$ are constants corresponding to constant values of flux through the Riemann surface.
Following \cite{Maldacena:2000mw,Gaiotto:2009gz,Benini:2009mz,Bah:2011vv,Bah:2012dg}, the flux terms can be seen as arising from making the following ansatz for the gauge potentials
\begin{eqnarray}
A^{(i)} &=& A^{(i)} + \kappa p_i\, \omega_{{}_\Sigma}\,, \qquad \qquad  \kappa = \pm 1,\nn \\
A^{(i)} &=& A^{(i)} + 4\pi p_i\, x_1 dx_2\,, \qquad \qquad \kappa = 0,
\end{eqnarray}
where $\omega_{{}_\Sigma}$ is the spin-connection on the Riemann surface which satisfies $d\omega_{{}_\Sigma} = \kappa vol_\Sigma.$ 

As in \cite{Maldacena:2000mw,Gaiotto:2009gz,Benini:2009mz,Bah:2011vv,Bah:2012dg}, in order to achieve a supersymmetric reduction the values of $p_1$ and $p_2$ are constrained such that $p_1 + p_2 = -\frac{\kappa}{2m},$ where $m$ is the seven-dimensional supergravity coupling constant. This is such that the gauge potential term cancels the spin connection contribution in the gravitino variation. When considering supersymmetric truncations we are then left with a one-parameter family of solutions coming from the difference, $p_1 - p_2 \equiv \frac{z}{2m},$  where $z$ is constrained to be a rational number \cite{Bah:2011vv,Bah:2012dg}.\footnote{Without loss of generality, we will assume $z \ge 0.$ The case for negative $z$ is equivalent to interchanging $\{A^{(1)}, \lambda_1\}$ with $\{A^{(2)}, \lambda_2\}$ at fixed $z.$}  We thus define
\begin{equation}\label{eq:psusy}
p_1 = -\frac{\kappa - z}{4m},\qquad p_2 = -\frac{\kappa + z}{4m},
\end{equation}
Note that for $\kappa \neq 0$ the conventions for $z$ chosen above differ from those in \cite{Bah:2011vv,Bah:2012dg} by the relation $z_{there} = -\kappa z_{here}.$ This is chosen in order to achieve a uniform notation which contains all three cases $\kappa = \pm 1,\, 0.$ One should also note that in terms of attaining a consistent truncation of the bosonic equations of motion the $p_i$ are completely arbitrary.\footnote{I would like to thank Nikolay Bobev for pointing this out and also for discussions on this point.} For the presentation of the bosonic reduction we will leave them arbitrary; however, for the discussion of a-maximization and the majority of the paper we will necessarily fix $p_1$ and $p_2$ to their supersymmetric values (\ref{eq:psusy}).

Moving on with the reduction, for the three-form we take
\begin{equation}\label{eq:c3ansatz}
S_3 = c_3 + c_1\wedge vol_\Sigma,
\end{equation}
where $c_3$ and $c_1$ are a three-form and a one-form in five-dimensions and $vol_\Sigma$ is the volume form on the Riemann surface.
$S_3$ satisfies an odd-dimensional self-duality equation in seven-dimensions (\ref{eq:U1eoms}) which implies the constraint
\begin{equation}\label{eq:c3constraint}
c_3 = -\frac{1}{m}e^{-\frac{8B}{3}+4\lambda_1+4\lambda_2} *_5\Big[dc_1 + \frac{2}{\sqrt{3}\,m}\left(p_1F^{(1)} + p_2 F^{(2)}\right) \Big],
\end{equation}
where $*_5$ is the Hodge star operator on the five-dimensional space-time. As we will see, this condition implies that $c_1$ has a mass term in the five-dimensional theory and does not correspond to a conserved current in the dual gauge theory. Finally, the scalars $\lambda_1$ and $\lambda_2$ are taken to be independent of the coordinates on the Riemann surface.

With this ansatz we can now reduce the seven-dimensional equations of motion. The details of the reduction of the equations of motion in the $U(1)^2$ truncation of the $\mathcal N = 4$ gauged supergravity can be found in appendix \ref{appsec:BosRed}. The reduced equations of motion can be derived from the following five dimensional Lagrangian
\begin{eqnarray}\label{eq:5dlag}
\mathcal L &=& R *_5 1 - 2  e^{\frac{4B}{3}-4\lambda_1} F^{(1)}\wedge*_5F^{(1)}- 2  e^{\frac{4B}{3}-4\lambda_2} F^{(2)}\wedge*_5F^{(2)} - 2 e^{-\frac{8B}{3}+4\lambda_1+4\lambda_2}F^{(0)}\wedge*_5F^{(0)} \nn \\
&& + 8 A^{(0)}\wedge F^{(1)}\wedge F^{(2)}  - 6m^2e^{-4B-4\lambda_1-4\lambda_2}c_1\wedge *_5 c_1 - \ft{10}{3} dB\wedge *_5dB  \nn \\
&& - 5d(\lambda_1+\lambda_2)\wedge*_5 d(\lambda_1+\lambda_2) - d(\lambda_1-\lambda_2)\wedge*_5 d(\lambda_1-\lambda_2) - V *_5 1,
\end{eqnarray}
where the potential is given by
\begin{eqnarray}\label{eq:5dV}
V &=& -2\kappa e^{-\frac{10B}{3}} - 4m^2 e^{-\frac{4B}{3}}\left(e^{2\lambda_1+2\lambda_2}+\fft12e^{-2\lambda_1-4\lambda_2} +\fft12e^{-4\lambda_1-2\lambda_2}-\fft18e^{-8\lambda_1-8\lambda_2}\right) \nn \\
&& + 2e^{-\frac{16B}{3}}\left(p_1^2e^{-4\lambda_1}+p_2^2e^{-4\lambda_2}\right),
\end{eqnarray}
and we have defined the gauge field
\begin{equation}
A^{(0)} = -\sqrt{3}\Big(c_1 - \fft{2}{\sqrt{3}\,m}(p_2A^{(1)} + p_1A^{(2)})\Big).
\end{equation}
For the supersymmetric values in (\ref{eq:psusy}), the class of solutions found in \cite{Bah:2011vv,Bah:2012dg} are critical points of $V$. For the explicit form of these solutions in the present notation see appendix \ref{appsec:ads}.

From now on we focus on the supersymmetric case with $p_1$ and $p_2$ given by (\ref{eq:psusy}). It would be interesting to investigate the potential for generic values of $p_1$ and $p_2$. Solutions not satisfying (\ref{eq:psusy}) would correspond to non-supersymmetric vacua. We leave this for future study.

The Lagrangian has the form required for five-dimensional $\mathcal N = 2$ gauged supergravity coupled to two vector multiplets and a single hypermultiplet. We will discuss the details of the matching to five-dimensional supergravity in the next section. It should be noted first that the action is not completely off-shell. In particular, there is only one hyperscalar showing up explicitly above. Another hyperscalar can be introduced as a Stuckelberg scalar for $c_1$ while two other hyperscalars have been fixed to constant values implicitly in the reduction.

Nevertheless, we have verified that this is a consistent truncation and the supersymmetric structure is left intact. In particular, the very special geometry of the vector multiplet sector is fully realized and this truncation contains the necessary information to determine the superconformal $U(1)_R$ and central charge at the $AdS_5$ critical point.

Before discussing the $\mathcal N = 2$ supergravity details of this reduction we would first like to comment on the spectrum of this theory about the critical point.

\subsubsection{An Aside on the Linearized Spectrum around the $AdS_5$ Critical Point}

One can perform a linearized analysis of this theory about the supersymmetric $AdS_5$ critical points (\ref{eq:AdSsol}) in order to determine the five-dimensional masses of the massive vector and scalars. Linearizing the equations of motion in section \ref{appsec:BosRed} we find the following spectrum.

For the vectors, we find one massive mode $c_1,$ with mass
\begin{equation}
m^2_{c_1}L^2 = 12 + |\kappa|\frac{4 + \kappa 4\sqrt{1 + 3 z^2}}{z^2},
\end{equation}
where the $AdS$ radius $L$ is found by evaluating the potential and equating $12/L^2 = - \tilde{V},$ where we use a tilde to represent quantities that have been evaluated on the solution (\ref{eq:AdSsol}). Using the relation $(\Delta-1)(\Delta-3)=m_v^2L^2$ between the mass-squared $m_v^2$ of the massive vector and the conformal dimension $\Delta$ of its dual operator one finds
\begin{equation}
\Delta_{c_1} = 2 + \fft1{z}\sqrt{4|\kappa|+13z^2 + 4\kappa\sqrt{1+3z^2}}.
\end{equation}

Additionally, there are two linear combinations of massless gauge fields which can be written
\begin{eqnarray}
A^{m=0}_1&=& (\kappa-z)e^{4\tilde\lambda_2} A^{(1)} - (\kappa+z)e^{4\tilde\lambda_1} A^{(2)}  \\
A^{m=0}_2&=& 2 \sqrt{3} m^2c_1 + \left( (\kappa+z) + \ft{2m^4}{\kappa+z}e^{4\tilde B-8\tilde\lambda_1-4\tilde\lambda_2}\right)A^{(1)}-\left((\kappa-z)+\ft{2m^4}{\kappa-z}e^{4\tilde B-4\tilde\lambda_1-8\tilde\lambda_2} \right)A^{(2)}. \nn
\end{eqnarray}
%
One linear combination of these two gauge fields corresponds to the graviphoton, which is considered dual to the $U(1)_R$ current, and the other belongs to a massless vector multiplet -- which is dual to a flavor $U(1)$ current. Once we find the $U(1)_R$ we will see that it will determine the linear combination corresponding to the graviphoton. This will be done by demonstrating that there is a consistent truncation to minimal $\mathcal N = 2$ gauged supergravity, whose bosonic sector contains only the metric and graviphoton -- which are dual to the stress-energy tensor and $R$-current, respectively.

For the scalars, we find the following diagonalized modes
\begin{eqnarray}
\chi_1 &=& 3\left(-\ft23 B + \lambda_1 + \lambda_2\right) - \frac{1}{z}\left(\kappa - \sqrt{|\kappa|+3z^2} \right)(\lambda_1 - \lambda_2), \\
\chi_{\pm} &=& \left(\pm 1 +\frac{1}{z} \big(4\kappa(\kappa+\sqrt{|\kappa|+3z^2})+13z^2\,\big)^{1/2}\right)\times \nn \\
&&\left(\lambda_1-\lambda_2 + \frac{1}{z}(\kappa-\sqrt{|\kappa|+3z^2}\,)\left(-\ft23 B + \lambda_1 + \lambda_2\right)\right) \nn \\
&& \mp \, \frac{8}{z} \sqrt{|\kappa|+3z^2}(B + \lambda_1 + \lambda_2),
\end{eqnarray}
with corresponding masses
\begin{eqnarray}
m_{\chi_1}^2 L^2  &=& -4 \nn \\
m_{\chi_\pm}^2 L^2 &=& 10 + \fft{4|\kappa|}{z^2} (1 + \kappa\sqrt{1 + 3 z^2}) \pm \fft2{z}(13 z^2 + 4|\kappa|(1 +\kappa \sqrt{1 + 3 z^2})\,)^{1/2}.
\end{eqnarray}
The existence of the scalar $\chi_1$ with $m_{\chi_1}^2L^2 = -4$ is in accordance with the expectation that this scalar belongs to a massless vector multiplet. For a scalar, the relation between the scaling dimension of its dual operator and the mass-squared of the scalar $m_s^2$ is $\Delta(\Delta-4) = m_s^2L^2,$ using this we see that $\Delta_{\chi_1} = 2$.

Finally, when expanding around the AdS vacuum we can think of the two modes $\chi_\pm$ as belonging (along with $c_1)$ to a massive vector multiplet. One can check that the dimensions are related by
\begin{equation}
\Delta_{\chi_+} = \Delta_{c_1} + 1 = \Delta_{\chi_-} + 2,
\end{equation}
as expected for the two neutral scalars and the vector in a massive vector multiplet in $AdS_5.$ Note that a complete massive vector multiplet also contains a charged scalar. In this truncation this has already been set to zero.

Here we have verified that, indeed, there are two massless $U(1)$ gauge bosons in the bulk and $a$-maximization is likely required in the dual theory to determine the superconformal $U(1)_R$. Next, we identify the required supergravity information of this reduction in order to proceed with the dual of $a$-maximization.

\subsection{Identifying the $\mathcal N = 2$ supergravity structure}

To proceed with the dual description of $a$-maximization we must first determine some details of the particular gauged supergravity in which this reduction belongs. The necessary information will be ascertained from the bosonic lagrangian (\ref{eq:5dlag}) and the fermion supersymmetry variations (\ref{eq:gravitino}) and (\ref{eq:spin1/2vars}). The details of the fermion reduction and supersymmetry transformations are given in appendix \ref{sec:N2vars}. Recall that the relevant information required consists of the $c_{IJK}$ coefficients, the gauge fields $A^I_\mu,$ the constrained scalars $X^I,$ the prepotentials $P_I$ and (since the one-form $c_1$ is Higgsed) the scalars in the hypermultiplet sector. 

First, the gauge fields are identified simply as $A_\mu^I \equiv 2 A_\mu^{(I)},$ for $I = 0,1,2.$ For the $c_{IJK}$ coefficients, there is only one nonvanishing component $c_{012} = 1.$ From the structure of the spin-$1/2$ variations in (\ref{eq:spin1/2vars}) we can determine that the only nontrivial hyperscalar is given by the linear combination $B+\lambda_1+\lambda_2.$ We can then read off the constrained scalars $X^I$ and the prepotentials $P_I$ from the gravitino variation (\ref{eq:gravitino}). The $X^I$ are given by
\begin{equation}
X^0 = e^{\frac{4B}{3}-2\lambda_1-2\lambda_2}, \qquad X^1 = e^{-\frac{2B}{3}+2\lambda_1}, \qquad X^2 = e^{-\frac{2B}{3}+2\lambda_2},
\end{equation}
the prepotentials are
\begin{equation}\label{eq:PI}
P_0 = \frac{m}{2}e^{-2B-2\lambda_1-2\lambda_2}, \qquad P_1 = m + \frac{\kappa + z}{4m}e^{-2B-2\lambda_1-2\lambda_2}, \qquad P_2 = m + \frac{\kappa - z}{4m}e^{-2B-2\lambda_1-2\lambda_2},
\end{equation}
and the superpotential is
\begin{equation}\label{eq:P}
P \equiv X^I P_I = m e^{-\frac{2B}{3}}\big(e^{2\lambda_1}+e^{2\lambda_2}+\ft12e^{-4\lambda_1-4\lambda_2}\big) + \fft1{4m}e^{-\frac{8B}{3}}\big((\kappa-z)e^{-2\lambda_1}+(\kappa+z)e^{-2\lambda_2}\big).
\end{equation}
Additionally, one can verify that this superpotential reproduces the scalar potential\footnote{Note that in the case of generic $p_1$ and $p_2$ this is no longer true.} in (\ref{eq:5dlag}) via the relation
\begin{equation}
V = 2 (\mathcal G^{-1})^{ij}\partial_i P\partial_j P - \fft43 P^2,
\end{equation}
where $(\mathcal G^{-1})^{ij}$ is the inverse of the scalar metric which can be read off from the Lagrangian (\ref{eq:5dlag}).

Before moving on we would like to note that, due to the mass term for $c_1$, there are two natural choices of basis for the very special geometry. We have chosen to work with a basis in which the kinetic terms of the gauge fields are diagonalized, i.e. so that the matrix $G_{IJ}$ is diagonal. This is at the cost of having nontrivial mass-like couplings between the gauge fields of the form $A^I\wedge *_5 A^J$ for all $I,J = 0,1,2$. An alternative basis corresponds to diagonalizing the mass term, in which case $c_1$ is chosen as one of the vectors of very special geometry, say $A^0.$ This has the advantage of uniquely identifying the massive mode with a single vector of very special geometry. This is important because in this case we are naturally lead to disregard the scalar $X^0$ in the R-charge since $X^0$ is paired with the massive gauge field $A^0 \equiv c_1$, which is dual to an anomalous current. This choice would lead to a significantly more complicated expression for the $X^I$ and the coefficients $c_{IJK}.$ In the end, both choices yield the same results for the R-charge (as they should) and, for sake of clarity, we choose to work with the simpler conventions in which the gauge kinetic terms are diagonal.

We are now ready to move on and discuss the dual of $a$-maximization within this reduction.

\section{$a$-maximization for wrapped M5$\textrm{s}$ from gauged supergravity}\label{sec:amax}

Moving forward with the procedure outlined in \cite{Tachikawa:2005tq} for implementing $a$-maximization within the supergravity description we should minimize (\ref{eq:P}) with respect to the vector scalars. However, since $c_1$ is massive (and does not correspond to a conserved current in the dual gauge theory) including it requires one to consider the dual of the more general formalism involving lagrange multipliers to enforce the anomaly free condition developed within gauge theory in \cite{Kutasov:2003ux}. As discussed previously the implementation of this was also determined in \cite{Tachikawa:2005tq}. In this case we separate the prepotentials as
\begin{equation}
P_I = P_{(0)\,I} + m^a_I P_a,
\end{equation}
with all dependence on the hypermatter existing in the $P_a.$ For the case at hand we have a single $P_a = e^{-2B-2\lambda_1-2\lambda_2}$ and the constants $P_{(0)\,I}$ and $m^a_I$ can be read off from (\ref{eq:PI}).

In this prescription $P_a$ are the analogue of the Lagrange multipliers, which are related to gauge and superpotential couplings in the field theory \cite{Barnes:2004jj,Kutasov:2004xu,Tachikawa:2005tq}. In fact, discussed in \cite{Tachikawa:2005tq}, for the case of a massive vector multiplet, $P_a$ is precisely dual to the gauge coupling.

$a$-maximization is still implemented by identifying the trial $a$-function with the inverse cube of the superpotential
\begin{equation}
a(s) \propto \frac{1}{P^3}.
\end{equation}
The maximization of this is then equivalent to minimization of $P$ with respect to all of the scalars in the theory. We then arrive at the following conditions (noting that $X^0 = (X^1X^2)^{-1}$)
%
%
%
\begin{eqnarray}\label{eq:Pext}
&&0=\frac{\delta P}{\delta X^1} = - \frac{m}{2}e^{-6\lambda_1-4\lambda_2} + m + \frac{\kappa+z}{4m}e^{-2B-2\lambda_1-2\lambda_2}, \nn\\
&&0=\frac{\delta P}{\delta X^2} = - \frac{m}{2}e^{-4\lambda_1-6\lambda_2} + m + \frac{\kappa-z}{4m}e^{-2B-2\lambda_1-2\lambda_2}, \nn\\
&&0=\frac{\delta P}{\delta P_a} = \frac{m}{2}e^{\frac{4B}{3}-2\lambda_1-2\lambda_2} + \frac{\kappa+z}{4m}e^{-\frac{2B}{3}+2\lambda_1} + \frac{\kappa-z}{4m}e^{-\frac{2B}{3}+2\lambda_2}.
\end{eqnarray}

By solving the first two expressions in (\ref{eq:Pext}) for $X^1$ and $X^2$, one can determine the superpotential as a function of $\lambda \equiv P_a$
\begin{equation}
P(\lambda) = 3(P_0 P_1 P_2)^{1/3}.
\end{equation}
From the holographic RG flow perspective one is tempted to treat $P(\lambda)$ as the superpotential that governs the BPS flow of domain wall solutions of the form
\begin{equation}\label{eq:domwall}
ds_5^2 = e^{2A(r)}dx_\mu dx^\mu + dr^2,
\end{equation}
with a varying hyperscalar satisfying
\begin{equation}
\frac{d\lambda}{dr} \propto \frac{dP(\lambda)}{d\lambda}.
\end{equation}
However, by solving the first two expressions in (\ref{eq:Pext}) we have implicitly assumed that $X^1$ and $X^2$ are constant.\footnote{By constant we mean independent of the radial direction in a holographic RG flow.} Solving for these in terms of $\lambda$ is only consistent if $\lambda$ is also constant. So it appears $P(\lambda)$ does not have a direct interpretation in terms of a holographic RG flow.

Nevertheless, in the correspondence $P(\lambda)$ should be analogous to the function $a(\lambda)$ introduced in \cite{Kutasov:2003ux}, which can be thought of as an $a$-function with argument $\lambda$ being related to the gauge coupling. In the case discussed here $a(\lambda) \propto P(\lambda)^{-3}$ simply tells you the possible values of the holographic central charge at the critical point, while allowing the value of $\lambda$ to vary. Analyzing the behavior of this function as we vary the argument from $\lambda = 0$ to $\lambda = \lambda_*,$ where $\lambda_*$ corresponds to the value at the critical point, we see that $a(\lambda) \propto P(\lambda)^{-3}$ reaches a maximum at $\lambda_*.$ This is in seeming accordance with the $a$-theorem \cite{Cardy:1988cwa,Komargodski:2011vj}. However, the interpretation of this isn't entirely clear, since the full set of BPS conditions are not satisfied for any point other than the $\lambda_*$ critical point. Thus, in order to make connections between this and any sort of holographic RG flow, it is necessary to analyze the complete set of coupled first order BPS equations. Representing the scalars in the theory as $\phi^i$ the BPS equations are of the form
\begin{eqnarray}
\partial_r A(r) & \propto & P \nn \\
\partial_r\phi^i &\propto & (\mathcal G^{-1})^{ij}\partial_j P,
\end{eqnarray}
where $A(r)$ is as in (\ref{eq:domwall}) and $\partial_j$ represents the derivative with respect to the scalar $\phi^j.$ Equivalent BPS equations were found in \cite{Bah:2012dg} and used to find holographic RG flows starting at the asymptotically $AdS_7$ solution in the UV and ending at the $AdS_5$ critical points in the IR. It would be interesting to further study these for any relevant deformations of the $AdS_5$ critical points.


Moving on, we first note that the conditions (\ref{eq:Pext}) are satisfied for the solutions (\ref{eq:AdSsol}).\footnote{In the following we fix $m=2$ as in \cite{Bah:2012dg}.} In fact, (\ref{eq:Pext}) are precisely the conditions required for the gaugino and hyperino supersymmetry variations (\ref{eq:spin1/2vars}) to vanish in the case of constant scalars and vanishing vectors.

Now we will identify the $U(1)_R$ for these solutions. Following \cite{Tachikawa:2005tq}, the coefficients of the superconformal $R$-charge are given by $s^1 = \tilde{X}^1/\tilde{P}$ and $s^2 = \tilde{X}^2/\tilde{P}.$\footnote{Recall that the vector field $c_1$ has been Higgsed and so $A^0$ should not be associated with a conserved current in the gauge theory. This can also be seen from the fact that $P_{(0)\,0} = 0$, which implies that $X^0$ only enters the $a$-function through the anomalous contribution (as it is multiplied by the lagrange multiplier) and so should not contribute to the superconformal $R$-charge.} Recalling that $A^1$ and $A^2$ correspond to the gauging of the $\phi_1$ and $\phi_2$ directions on the $S^4$, respectively, we see that the $R$-charge should be identified with the killing vector\footnote{The factor of $2$ originates from the normalization of $A^{(I)}$ in (\ref{eq:metricS4}) and the relation $A^I = 2A^{(I)}.$}
\begin{equation}
2\left(s^1 \partial_{\phi_1} + s^2 \partial_{\phi_2}\right).
\end{equation}
Evaluating this we find
\begin{equation}
\ft12[(\partial_{\phi_1} + \partial_{\phi_2}) + \epsilon(\partial_{\phi_1} - \partial_{\phi_2})],
\end{equation}
where we have defined
\begin{equation}
\epsilon = -\frac{\kappa+\sqrt{|\kappa|+3z^2}}{3z}.
\end{equation}
Up to an overall normalization\footnote{Also recall the relation $z_{here} = -\kappa z_{there}$ for $\kappa = \pm 1$ when comparing with \cite{Bah:2011vv,Bah:2012dg}.}, this choice of $U(1)_R$ reproduces the result of \cite{Bah:2012dg} and provides a nontrivial check of the procedure outlined in \cite{Tachikawa:2005tq}. Furthermore, this demonstrates that the procedure in \cite{Tachikawa:2005tq} can be used to identify the $U(1)$ isometry which is dual to the $U(1)_R$ in a string/M-theory reduction.

We can also demonstrate that $(s^1\partial_{\phi_1} - s^2\partial_{\phi_2})$ corresponds to the left-over non-$R$ $U(1)$ flavor symmetry. Indeed, when evaluated at the critical point
\begin{eqnarray}
s^1\tilde{P}_1 - s^2 \tilde{P}_2 =0.
\end{eqnarray}
This implies that this $U(1)$ acts trivially on the fermionic coordinates.

A few comments about the preceding results are in order. For $\kappa = -1$ there are two special values of $z,$ given by $z=0$ and $z=1,$ for which $a$-maximization is not required in the gauge theory.\footnote{I would like to thank Brian Wecht for pointing this out.} There should be an analogous statement in supergravity. These choices of $z$ correspond to the $\mathcal N = 1$ and $\mathcal N = 2$ Maldacena-N\'{u}\~{n}ez (MN) solutions \cite{Maldacena:2000mw} for $z=0$ and $z=1$ respectively. The case $z = 0$ can be seen directly by noticing that in this case there is a symmetry that interchanges $\{A^{1}_\mu,X^1\}$ with $\{A^{2}_\mu,X^2\}.$ Thus for $z=0$ we see that $X^1 = X^2$ and the $U(1)_R$ is simply dual to $(\partial_{\phi_1} + \partial_{\phi_2})$ and can be determined without considering the minimization of $P.$  Note also that the flavor $U(1)$ isometry identified above becomes simply $(\partial_{\phi_1} - \partial_{\phi_2})$ in this case. The case of $z = 1$ is more intricate. The point is that for $z = 1$ the dual SCFT has $\mathcal N = 2$ supersymmetry in four dimensions. In this case there should exist an extended consistent truncation in which, for example, the $A^{2}_\mu$ gauge group is enhanced from $U(1)$ to $SU(2).$ This is such that the supergravity has total gauge group $SU(2)\times U(1)$ corresponding to the $SU(2)\times U(1)$ R-symmetry of an $\mathcal N = 2$ gauge theory. This is possible because for $z=1$ the seven dimensional $A^{(2)}_M$ does not have any components along the Riemann surface and it is natural to decompose the gauge group as $SO(5) \rightarrow SO(4) \cong SU(2)\times SU(2) \rightarrow SU(2)\times U(1)$ in the reduction, instead of the  $U(1)\times U(1)$ discussed here. In fact, such a reduction to a five-dimensional $SU(2)\times U(1)$ gauged supergravity directly from M-theory containing the $\mathcal N = 2$ MN solution was discussed in \cite{Gauntlett:2007sm}.

\subsubsection{Truncation to minimal supergravity}

Another interesting point is that, having identified the $U(1)_R,$ it is straightforward to show that there exists a truncation to five-dimensional minimal gauged supergravity. This is natural since the field content of minimal supergravity is dual precisely to the superconformal stress-enegy tensor multiplet in which the $R$-current sits. However, the choice of graviphoton in the truncation is difficult to see {\it a priori} due to the nontrivial value of the scalars at the critical point (\ref{eq:AdSsol}). But utilizing the $a$-maximization results, the truncation proceeds directly by fixing all of the scalars at the critical point (\ref{eq:AdSsol}) and making the following identifications on the gauge fields
\begin{eqnarray}
0&=& c_1,  \nn \\
0&=& \tilde{X}^1 A_1 - \tilde{X}^2 A_2, \nn \\
A_R &=& \frac{1}{2}(\tilde{X}^1 A_1 + \tilde{X}^2 A_2),
\end{eqnarray}
where the gauge fields with lower indices are $A_I \equiv G_{IJ}A^J$, with $G_{IJ}$ given in (\ref{eq:vecscmetric}). According to the discussion above these identifications amount to setting both the massive gauge field $c_1$ and the flavor gauge field $(\tilde{X}^1A_1 - \tilde{X}^2 A_2)$ to zero while retaining only the gauge field dual to the $U(1)_R$, defined as $A_R$ above. This can be shown to be a consistent truncation. Furthermore, with these identifications, the lagrangian (\ref{eq:5dlag}) reduces to simply
\begin{eqnarray}
\mathcal L &=& R*_51 + \frac{12}{L^2}*_51 - \fft32 F_R\wedge *_5 F_R  + A_R\wedge F_R \wedge F_R,
\end{eqnarray}
where $12/L^2 = - \tilde{V}$. This is the lagrangian for minimal gauged supergravity in five dimensions. So we see that knowledge of the relation to $a$-maximization makes it straightforward to determine the $A_R$ in the truncation to the minimal gauged supergravity sector, which is a universal truncation for reductions with supersymmetric $AdS_5$ critical points \cite{Gauntlett:2006ai}.

\subsubsection{The Holographic Central Charge}

Finally, we can evaluate the holographic central charge $a \propto P^{-3}.$ To fix the normalization we compare to the holographic Weyl anomaly \cite{Henningson:1998gx}. Using the result of \cite{Maldacena:2000mw} we have
\begin{equation}
a = \frac{8(1-g)}{3\kappa}L^3N^3,
\end{equation}
where $L = 3\tilde{P}^{-1}$ is the $AdS_5$ radius, which can extracted from the gravitino variation in the $AdS$ vacuum or simply from evaluating the potential (\ref{eq:5dV}). Also for the torus one should set $\frac{(1-g)}{\kappa}$ to one. Evaluating this we find
\begin{equation}
a = \frac{(1-g)}{\kappa}N^3\left(\frac{\kappa - 9\kappa z^2 + (|\kappa|+3z^2)^{3/2}}{48z^2}\right),
\end{equation}
which agrees precisely with \cite{Bah:2012dg}.

\section{Discussion}\label{sec:conc}

\subsection{Relation to the Holographic $c$-theorem}

The $a$-function discussed in this paper is also naturally seen to arise in the context of the holographic $c$-theorem \cite{Girardello:1998pd,Freedman:1999gp,Myers:2010xs,Myers:2010tj,Liu:2010xc}. In the holographic $c$-theorem\footnote{The holographic $c$-theorem has been analyzed in arbitrary dimensions, in the following we restrict to the case of a five-dimensional bulk. Additionally, $a = c$ at leading order in large $N$ and so the holographic $c$-theorem is also an $a$-theorem.} one considers domain wall solutions of the form
\begin{equation}
ds_5^2 = e^{2A(r)}dx_\mu dx^\mu + dr^2,
\end{equation}
and defines the $a$-function as
\begin{equation}
a(r) = \frac{\pi^2}{\kappa^2} \frac{1}{(A'(r))^3}.
\end{equation}

It follows from the null-energy condition that this function generically decreases along holographic RG flows, i.e. for decreasing $r$. It is standard knowledge that domain wall solutions in supergravity often admit a first order formalism in terms of the superpotential $P$, in which the warp factor satisfies
\begin{equation}
A'(r) = \frac{1}{3} P.
\end{equation}
Using this we see directly that the $a$-function defined in studies of the holographic $c$-theorem is proportional to $P^{-3}.$ This is interesting because we see that both the holographic $c$-theorem and the dual of $a$-maximization utilize precisely the same $a$-function in the holographic context. Furthermore, it would be interesting to make a connection between the $a$-function discussed here and that recently constructed in a proof of the a-theorem \cite{Komargodski:2011vj} and also considered in various holographic contexts in \cite{Elvang:2012st,Hoyos:2012xc}.

\subsection{Potential Future Directions}

We have provided evidence for the supergravity description of $a$-maximization presented in \cite{Tachikawa:2005tq}. This allows one to determine the appropriate $U(1)$ isometry which is dual to the superconformal $U(1)_R,$ in situations where a precise gauged supergravity dual is known. We end with a brief discussion of possible directions for further study.

For $AdS_5$ solutions of type IIB \cite{Gabella:2009ni} (as well as in $AdS_4$ solutions in M-theory \cite{Gabella:2011sg}) the $U(1)_R$ is dual to the Reeb vector and is associated to a certain contact structure on the internal manifold. It would be very interesting if the $U(1)_R$ determined for the $M$-theory reduction considered here has a similar geometric interpretation.

Recently, a proposal utilizing a form of $a$-maximization with inequalities was discussed in \cite{Hook:2012fp} and used to detect emergent global symmetries in gauge theory. It would be interesting if this procedure also has a dual description in supergravity.

Finally, it would be interesting to extend these results on $a$-functions to other dimensions. For example, in $AdS_4$ solutions of four-dimensional gauged supergravities perhaps contact with the procedure of $F$-maximization \cite{Jafferis:2010un}, which also has an extension along RG flows \cite{Amariti:2011xp}, and also newly found anomalies in $\mathcal N = 2$ SCFTs in three-dimensional gauge theories \cite{Closset:2012vg,Closset:2012vp} could potentially be made. There has also been recent work in attempts to understand the existence of an $a$-theorem in six-dimensions \cite{Elvang:2012st}, perhaps similar constructions to that discussed here could be used to gain insight into whether or not a six-dimensional $a$-theorem holds.

\section*{Acknowledgements}

The author would like to thank Ibrahima Bah, Nikolay Bobev, Jim Liu and Brian Wecht for useful discussions and also for comments on a draft of this paper. The author would also like to thank the Simons Center for Geometry and Physics for hospitality during the 2012 Summer Simons Workshop in Mathematics and Physics where part of this work was completed. This work was supported by the U.S. Department of Energy under Grant No. DEFG02- 97ER41027.

\appendix

\section{Consistent truncation of $\mathcal N = 4$ gauged supergravity on a Riemann Surface}\label{app:N4trunc}

Here we present a consistent truncation of $\mathcal N = 4$ $SO(5)$ gauged supergravity in seven dimensions to a five dimensional gauged supergravity by reducing the seven dimensional theory on a Riemann surface. The resulting theory has the structure of five dimensional $\mathcal N = 2$ supergravity coupled to two vector multiplets and a single hypermultiplet, however, as we will see, the truncation does not fill out the hypermultiplet.

\subsection{$U(1)^2$ Truncation of seven-dimensional $\mathcal N = 4$ $SO(5)$ gauged supergravity}

We begin by describing a truncation of seven-dimensional $\mathcal N = 4$ $SO(5)$ gauged supergravity to the $U(1)^2$ sector, corresponding to the Cartan of $SO(5).$ For the most part, this truncation was worked out in \cite{Liu:1999ai}, here we present the details. In particular, we include the explicit form of the gauge field Chern-Simons term in the gauge field equation of motion. This was not explicitly given in \cite{Liu:1999ai} as they were interested in electric gauge field configurations for which this term vanishes. We follow the conventions in \cite{Liu:1999ai} but with the gauge coupling in the $U(1)^2$ truncation set to $g = 2m.$

The details of the $\mathcal N = 4$ theory are reviewed in \cite{Liu:1999ai} and a nice presentation of the complete $\mathcal N = 4$ equations of motion is given in \cite{Cvetic:2000ah,Donos:2010ax}. The bosonic field content of the $\mathcal N = 4$ theory consists of $SO(5)_g$ gauge fields $A^{ij},$ $(i,j = 1,\ldots,5)$ a real three form $S_3^i$ transforming in the $\bf{5}$ of $SO(5)_g$ and a symmetric unimodular matrix of scalars $T^{ij}$ parametrizing the coset $SL(5,\mathbb R)/SO(5)_c,$ where we have used subscripts $g$ and $c$ to distinguish the $SO(5)_g$ gauge group from the $SO(5)_c$ compensator in the coset. The truncation most straightforwardly follows by choosing a gauge in which $SO(5)_g$ is identified with $SO(5)_c$ and making the following ansatz on the fields,
\begin{eqnarray}
A^{(1)} &\equiv& A^{12} = -A^{21}, \nn \\
A^{(2)} &\equiv& A^{34} = -A^{43}, \nn \\
S_3 &\equiv& S_3^5,
\end{eqnarray}
and choosing a diagonal scalar matrix, with
\begin{eqnarray}
e^{2\lambda_1} \equiv T_{11} = T_{22}  \nn \\
e^{2\lambda_2} \equiv T_{33} = T_{44}  \nn \\
e^{-4\lambda_1-4\lambda_2} \equiv T_{55} ,
\end{eqnarray}
and all remaining terms set to zero.

Given this truncation, the equations of motion of the $\mathcal N = 4$ theory reduce to the following set of equations
\begin{eqnarray}\label{eq:U1eoms}
dS_3 &=& m e^{-4\lambda_1-4\lambda_2} *_7\!S_3 + \frac{2}{\sqrt{3}m}F^{(1)} \wedge F^{(2)}, \nn \\
d(e^{-4\lambda_1}*_7F^{(1)}) &=& - 2\sqrt{3} F^2\wedge dS_3 + \fft4{m} F^{(1)} \wedge F^{(2)} \wedge F^{(2)}, \nn \\
d(e^{-4\lambda_2}*_7F^{(2)}) &=& - 2\sqrt{3} F^1\wedge dS_3 + \fft4{m} F^{(2)} \wedge F^{(1)} \wedge F^{(1)}, \nn \\
d\left(*_7 d(3\lambda_1 + 2\lambda_2)\right) &=& -2 e^{-4\lambda_1} F^{(1)}\wedge*_7F^{(1)} - 6m^2 e^{-4\lambda_1-4\lambda_2} S_3\wedge *_7S_3 - m^2\frac{\delta V_7}{\delta \lambda_1} *_7\!1, \nn \\
d\left(*_7 d(2\lambda_1 + 3\lambda_2)\right) &=& - 2 e^{-4\lambda_2} F^{(2)}\wedge*_7F^{(2)} - 6m^2 e^{-4\lambda_1-4\lambda_2} S_3\wedge *_7S_3 - m^2\frac{\delta V_7}{\delta \lambda_2} *_7\!1, \nn \\
R_{MN} &=& 5\partial_M(\lambda_1+\lambda_2)\partial_N(\lambda_1+\lambda_2) + \partial_M(\lambda_1-\lambda_2)\partial_N(\lambda_1-\lambda_2) \nn\\&& + 2e^{-4\lambda_1}F^{(1)}_{MP}F^{(1)}_{\,\,\,\,N}{}^{P} + 2e^{-4\lambda_2}F^{(2)}_{MP}F^{(2)}_{\,\,\,\,N}{}^{P} \nn\\
&& + 3m^2 e^{-4\lambda_1-4\lambda_2} S_{MPQ}S_{N}{}^{PQ} + \ft{1}{10}g_{MN}X,
\end{eqnarray}
where $V_7$ and $X$ are
\begin{eqnarray}
V_7 &=& e^{2\lambda_1+2\lambda_2} + \fft12 e^{-2\lambda_1-4\lambda_2} +  \fft12 e^{-4\lambda_1-2\lambda_2} - \fft1{16}e^{-8\lambda_1-8\lambda_2},  \\
X &=& -2 e^{-4\lambda_1}F^{(1)}_{MN}F^{(1)}{}^{MN} -2 e^{-4\lambda_2}F^{(2)}_{MN}F^{(2)}{}^{MN} - 4m^2e^{-4\lambda_1-4\lambda_2}S_{MNP}S^{MNP} - 16m^2 V_7. \nn
\end{eqnarray}

We will also consider the fermion supersymmetry variations in this truncation. These were also worked out in \cite{Liu:1999ai}. The fermions of the $\mathcal N = 4$ theory are a spin-$3/2$ gravitino $\psi_M$ and spin-$1/2$ fermions $\zeta_i$ transforming in the $\bf 4$ and $\bf 5$ representations of $SO(5)_c$ respectively.\footnote{Note that we have relabeled the spin-$1/2$ fermions $\zeta_1$ and $\zeta_2$ relative to \cite{Liu:1999ai}, where they are denoted by $\lambda_i$ in order to avoid confusion with the scalars $\lambda_1$ and $\lambda_2.$} After imposing the bosonic truncation above, the truncation of the fermions follows by identifying
\begin{equation}
\zeta_1 = \zeta_2, \qquad \zeta_3 = \zeta_4, \qquad \zeta_5 = -2\zeta_1 - 2\zeta_3,
\end{equation}
the third of which is consistent with the gamma-tracelessness condition $\Gamma^i\zeta_i = 0,$ where $\Gamma^i$ are gamma matrices of $SO(5)_c.$ Defining the linear combinations,
\begin{equation}
\tilde\psi_M = \psi_M + \ft12 \tilde{\gamma}_M\Gamma^5\lambda_5, \qquad \zeta^1 = \ft32 \Gamma^1\zeta_1 + \Gamma^3\zeta_3, \qquad \zeta^2 = \Gamma^1\zeta_1 + \ft32\Gamma^3\zeta_3,
\end{equation}
the fermion variations can be written
\begin{eqnarray}\label{eq:U1sqvars}
\delta\tilde\psi_M &=& \big[\nabla_M + m(A_M^{(1)}\Gamma^{12}+A_M^{(2)}\Gamma^{34}) + \ft{m}4e^{-4\lambda_1-4\lambda_2}\tilde{\gamma}_M + \ft12 \tilde{\gamma}_M\tilde{\gamma}^N\partial_N(\lambda_1+\lambda_2)  \nn \\
&& \kern2em + \ft12\tilde{\gamma}^N(e^{-2\lambda_1}F^{(1)}_{MN}\Gamma^{12}+e^{-2\lambda_2}F^{(2)}_{MN}\Gamma^{34}) - \ft{m\sqrt{3}}4\tilde{\gamma}^{NP}e^{-2\lambda_1-2\lambda_2} S_{MNP}\Gamma^5\big]\epsilon, \nn \\
\delta\zeta^1 &=& \big[\ft{m}4(e^{2\lambda_1}-e^{-4\lambda_1-4\lambda_2})-\ft14\tilde{\gamma}^M\partial_M(3\lambda_1+2\lambda_2) \nn \\
&& \kern2em - \ft18\tilde{\gamma}^{MN}e^{-2\lambda_1}F^{(1)}_{MN}\Gamma^{12} + \ft{m}{8\sqrt{3}}\tilde{\gamma}^{MNP}e^{-2\lambda_1-2\lambda_2}S_{MNP}\Gamma^5\big]\epsilon, \nn\\
\delta\zeta^2 &=& \big[\ft{m}4(e^{2\lambda_2}-e^{-4\lambda_1-4\lambda_2})-\ft14\tilde{\gamma}^M\partial_M(2\lambda_1+3\lambda_2)  \nn \\
&& \kern2em - \ft18\tilde{\gamma}^{MN}e^{-2\lambda_2}F^{(2)}_{MN}\Gamma^{34} + \ft{m}{8\sqrt{3}}\tilde{\gamma}^{MNP}e^{-2\lambda_1-2\lambda_2}S_{MNP}\Gamma^5\big]\epsilon,
\end{eqnarray}
where we denote the seven-dimensional gamma matrices with a tilde $\tilde\gamma_M.$

\subsection{Reduction on a Riemann Surface}\label{appsec:BosRed}

We can now demonstrate the consistency of the ansatz described in equations (\ref{eq:metricansatz})-(\ref{eq:c3ansatz}). For the three-form equation we find
\begin{eqnarray}
dc_3 &=& m e^{-4B-4\lambda_1-4\lambda_2} *_5c_1 + \frac{2}{m\sqrt{3}} F^{(1)}\wedge F^{(2)}, \nn \\
dc_1 &=& m e^{\ft{8B}3 - 4\lambda_1 -4\lambda_2} *_5c_3 + \frac{2}{m\sqrt{3}}(p_1F^{(2)} + p_2 F^{(1)}),
\end{eqnarray}
we impose the second equation above as a constraint and solve for $c_3$, which yields (\ref{eq:c3constraint}).

For the gauge fields we have
\begin{eqnarray}
d(e^{\ft{4B}3 - 4\lambda_1} *_5F^{(1)}) &=& -2\sqrt{3}p_2 dc_3 - 2\sqrt{3} F^{(2)}\wedge dc_1 \nn \\
&& + \frac{4}{m}\left(2p_2F^{(1)}\wedge F^{(2)} + p_1F^{(2)}\wedge F^{(2)} \right) \nn \\
d(e^{\ft{4B}3 - 4\lambda_2} *_5F^{(2)}) &=& -2\sqrt{3}p_1 dc_3 - 2\sqrt{3} F^{(1)}\wedge dc_1 \nn \\
&& + \frac{4}{m}\left(2p_1F^{(1)}\wedge F^{(2)} + p_2F^{(1)}\wedge F^{(1)} \right).
\end{eqnarray}

Combining these with the constraint (\ref{eq:c3constraint}) and the remaining equation from the three-form we get equations of motion for three coupled one-forms. Furthermore, the kinetic terms for the one-forms can be diagonalized by defining
\begin{equation}
A^{(0)} \equiv - \sqrt{3}c_1 + \frac{2}{m}\left(p_2A^{(1)}+p_1A^{(2)}\right).
\end{equation}

Reducing the $\lambda_1$ and $\lambda_2$ equations of motion yields,
\begin{eqnarray}
d(*_5d(3\lambda_1+2\lambda_2)) &=&- 2 e^{\frac{4B}{3}-4\lambda_1} F^{(1)}\wedge*_5F^{(1)} - 6m^2e^{\frac{8B}{3}-4\lambda_1-4\lambda_2}c_3\wedge*_5c_3 \nn \\
&&- 6m^2e^{-4B-4\lambda_1-4\lambda_2}c_1\wedge*_5c_1 - (2p_1^2 e^{-\frac{16B}{3}-4\lambda_1} + m^2e^{-\frac{4B}{3}}\frac{\delta V_7}{\delta\lambda_1})*_5 1, \nn \\
d(*_5d(2\lambda_1+3\lambda_2)) &=& -2 e^{\frac{4B}{3}-4\lambda_2} F^{(2)}\wedge*_5F^{(2)} - 6m^2e^{\frac{8B}{3}-4\lambda_1-4\lambda_2}c_3\wedge*_5c_3 \nn \\
&& - 6m^2e^{-4B-4\lambda_1-4\lambda_2}c_1\wedge*_5c_1 - (2p_1^2 e^{-\frac{16B}{3}-4\lambda_1} +  m^2e^{-\frac{4B}{3}}\frac{\delta V_7}{\delta\lambda_2}).
\end{eqnarray}

Finally we need to consider the reduction of the Einstein equation. The reduction of the seven-dimensional Ricci tensor is
\begin{eqnarray}
R^7_{\alpha\beta} &=& e^{\frac{4B}{3}}\left[R^5_{\alpha\beta} + \frac{2}{3} \nabla^\gamma\partial_\gamma B \eta_{\alpha\beta} - \frac{10}{3}\partial_\alpha B \partial_\beta B\right], \nn \\
R^7_{66} = R^7_{77} &=& - e^{\frac{4B}{3}} \nabla^\alpha\partial_\alpha B + \kappa e^{-2B},
\end{eqnarray}
with all other terms vanishing. The nonvanishing components of the Einstein equation then read
\begin{eqnarray}\label{eq:Beom}
\nabla^\alpha\partial_\alpha B &=&  \ft15 e^{\frac{4B}{3}-4\lambda_1}F^{(1)}_{\alpha\beta}F^{(1)\,\alpha\beta} + \ft15 e^{\frac{4B}{3}-4\lambda_2}F^{(2)}_{\alpha\beta}F^{(2)\,\alpha\beta} - \ft25e^{-\frac{8B}{3}+4\lambda_1+4\lambda_2}F^{(0)}_{\alpha\beta}F^{(0)\,\alpha\beta} \nn \\
&& - \ft{18}5m^2 e^{-4B-4\lambda_1-4\lambda_2} c_\alpha c^\alpha +\kappa e^{-\frac{10B}{3}} - \ft{8}{5}e^{\frac{16B}{3}}(p_1^2e^{-4\lambda_1} + p_2^2e^{-4\lambda_2}) \nn \\
&& + \ft45 m^2 e^{-\frac{4B}{3}}(e^{2\lambda_1+2\lambda_2} + \ft12 e^{-2\lambda_1-4\lambda_2}+ \ft12 e^{-4\lambda_1-2\lambda_2} - \ft18 e^{-8\lambda_1-8\lambda_2}) \\
R_{\alpha\beta}&=&   \ft{10}3 \partial_\alpha B \partial_\beta B + 5\partial_\alpha(\lambda_1+\lambda_2)\partial_\beta(\lambda_1+\lambda_2) + \partial_\alpha(\lambda_1-\lambda_2)\partial_\beta(\lambda_1-\lambda_2) \nn \\
&& 2e^{\frac{4B}{3}-4\lambda_1} F^{(1)}_{\alpha\gamma}F_\beta^{(1)\,\gamma} + 2e^{\frac{4B}{3}-4\lambda_2} F^{(2)}_{\alpha\gamma}F_\beta^{(2)\,\gamma} + 2e^{-\frac{8B}{3}+4\lambda_1+4\lambda_2} F^{(0)}_{\alpha\gamma}F_\beta^{(0)\,\gamma} \nn \\
&& + 6m^2 e^{-4B-4\lambda_1-4\lambda_2} c_\alpha c_\beta - (F^{(0)}_{\gamma\delta}F^{(0)\,\gamma\delta} + \ft23 \hat B  - \ft1{10} X^5 ) \eta_{\alpha\beta},
\end{eqnarray}
where $\hat B$ is given by the RHS of (\ref{eq:Beom}) and $X^5$ is
\begin{eqnarray}
X^5&=& - 2 e^{\frac{4B}{3}-4\lambda_1}F^{(1)}_{\alpha\beta}F^{(1)\,\alpha\beta} - 2  e^{\frac{4B}{3}-4\lambda_2}F^{(2)}_{\alpha\beta}F^{(2)\,\alpha\beta} + 4 e^{-\frac{8B}{3}+4\lambda_1+4\lambda_2}F^{(0)}_{\alpha\beta}F^{(0)\,\alpha\beta}\nn\\
&&  - 24m^2e^{-4B-4\lambda_1-4\lambda_2}c_\alpha c^\alpha  - 4 e^{\frac{16B}{3}}(p_1^2e^{-4\lambda_1} + p_2^2e^{-4\lambda_2}) \nn \\
&& - 16 m^2 e^{-\frac{4B}{3}}(e^{2\lambda_1+2\lambda_2} + \ft12 e^{-2\lambda_1-4\lambda_2} +  \ft12 e^{-4\lambda_1-2\lambda_2} - \ft1{8}e^{-8\lambda_1-8\lambda_2}).
\end{eqnarray}
The above equations of motion can be derived from the Lagrangian (\ref{eq:5dlag}).
\subsection{Supersymmetry transformations}\label{sec:N2vars}

We now turn to the reduction of the seven dimensional supersymmetry variations of the $U(1)^2$ truncation in (\ref{eq:U1sqvars}). The fermions include the gravitino $\psi_\alpha$ and two spin-$1/2$ particles $\zeta^1$ and $\zeta^2.$

We reduce the seven-dimensional spinors as\footnote{Our spinor and gamma matrix conventions can be found in appendix \ref{sec:gammaconv}.}
\begin{equation}\label{eq:fermred}
\psi_\alpha = e^{\frac{B}{3}} \psi_\alpha \otimes \eta\,, \qquad \psi_a = e^{\frac{B}{3}} \psi \otimes \gamma_a \eta\,, \qquad \zeta^i = e^{\frac{B}{3}} \zeta^i \otimes \eta\,,
\end{equation}
where on the right hand side $\psi_\alpha$ is a spin-$3/2$ fermion in five dimensions, $\psi$ and $\zeta_i$ are spin-$1/2$ fermions in five-dimensions and we use a Dirac spinor notation for all five-dimensional spinors. Also $\alpha = 0,\ldots,4$ and $a = 6,7$ are flat indices in seven-dimensions and $\eta$ is a constant spinor on the Riemann surface. For the supersymmetry parameter we take
\begin{equation}
\epsilon = e^{-\frac{B}{3}}\varepsilon \otimes \eta.
\end{equation}

Counting supercharges, the $\Gamma^{12}$ and $\Gamma^{34}$ projections in (\ref{eq:proj}) can be implemented by noting that the supersymmetry parameter $\epsilon$ in the $\mathcal N = 4$ theory is in the $\bf{4}$ of $USp(4)$ and can be decomposed into two $USp(2)$ spinors in the $\bf{2}$ representation. These $USp(2)$ spinors are then identified with each other. The choice of $\eta$ in the reduction further imposes one more constraint, implemented by the $\gamma^{67}$ projection. Overall this reduces the number of supercharges from thirty-two in the $\mathcal N = 4$ theory in seven dimensions to eight in the five-dimensional theory, which is appropriate for $\mathcal N = 2$ supergravity in five dimensions.

The reduction of the seven-dimensional supersymmetry variations is straightforward. The reduced five-dimensional gravitino variation becomes

\begin{eqnarray}\label{eq:gravitino}
\delta \hat\psi_\alpha &=& \Big[\nabla_\alpha - i m( A^{(1)}_\alpha+ A^{(2)}_\alpha + \ft{\sqrt{3}}{2}e^{-2B-2\lambda_1-2\lambda_2} c_\alpha) \nn \\
&& + \ft{i}{12}(\gamma_\alpha{}^{\beta\gamma}-2\delta_\alpha^\beta\gamma^\gamma)\big(e^{\frac{2B}{3}-2\lambda_1} F^{(1)}_{\beta\gamma} + e^{\frac{2B}{3}-2\lambda_2} F^{2}_{\beta\gamma} + e^{-\frac{4B}{3}+2\lambda_1+2\lambda_2}F^{(0)}_{\beta\gamma} \big) \\
&& +\Big(\ft{m}6e^{-\frac{2B}{3}}(e^{2\lambda_1}+e^{2\lambda_2}+\ft12e^{-4\lambda_1-4\lambda_2}) + \ft1{24m}e^{-\frac{8B}{3}}\big((\kappa-z)e^{-2\lambda_1}+(\kappa+z)e^{-2\lambda_2}\big)\Big)\gamma_\alpha\Big]\epsilon, \nn
\end{eqnarray}
where we have defined the shifted gravitino $\hat\psi_\alpha = \psi_\alpha + \frac{2}{3}\gamma_\alpha(\zeta^1+\zeta^2+\psi)$ and the field strength
\begin{equation}\label{eq:F0}
F^{(0)} = -\sqrt{3}\left(dc_1 + \fft{1}{2\sqrt{3}\,m^2}((\kappa+z)F^{(1)} + (\kappa-z)F^{(2)})\right).
\end{equation}
Additionally, the spin-$1/2$ variations can be put into the form:
\begin{eqnarray}\label{eq:spin1/2vars}
\delta \zeta^-&=& \Big[ -\ft14 \gamma^\alpha\partial_\alpha(\lambda_1-\lambda_2) + \ft{i}8\gamma^{\alpha\beta}(e^{\frac{2B}{3}-2\lambda_1}F^{(1)}_{\alpha\beta} - e^{\frac{2B}{3}-2\lambda_2}F^{(2)}_{\alpha\beta}) \nn\\
&& + \ft{m}4e^{-\frac{2B}{3}}\big(e^{2\lambda_1}-e^{2\lambda_2}\big) - \ft1{16m}e^{-\frac{8B}{3}}\big((\kappa-z)e^{-2\lambda_1} - (\kappa+z)e^{-2\lambda_1}\big)\Big]\epsilon \label{eqn:v1} \\
\delta\zeta^+ &=& \Big[ \ft12 \gamma^\alpha\partial_\alpha(B - \ft32\lambda_1- \ft32\lambda_2) + \ft{i}8\gamma^{\alpha\beta}(e^{\frac{2B}{3}-2\lambda_1}F^{(1)}_{\alpha\beta} + e^{\frac{2B}{3}-2\lambda_2}F^{(2)}_{\alpha\beta} - 2 e^{-\frac{4B}{3}+2\lambda_1+2\lambda_2} F^{(0)}_{\alpha\beta}) \nn\\
&& + \ft{m}4e^{-\frac{2B}{3}}\big(e^{2\lambda_1}+e^{2\lambda_2} - e^{-4\lambda_1-4\lambda_2}\big) + \ft1{16m}e^{-\frac{8B}{3}}\big((\kappa-z)e^{-2\lambda_1} + (\kappa+z)e^{-2\lambda_1}\big)\Big]\epsilon \label{eqn:v2}\\
\delta\psi&=& \Big[ \ft12 \gamma^\alpha\partial_\alpha(B + \lambda_1 + \lambda_2) - \ft{im\sqrt{3}}{2}e^{-2B-2\lambda_1-2\lambda_2}\gamma^\alpha c_\alpha + \ft{m}4 e^{-\frac{2B}{3}-4\lambda_1-4\lambda_2} \nn \\
&&  + \ft{1}{8m}e^{-\frac{8B}{3}}\big((\kappa-z)e^{-2\lambda_1} + (\kappa+z)e^{-2\lambda_2}\big)\Big]\epsilon\,, \label{eqn:h}
\end{eqnarray}
where we have defined $\zeta^- \equiv \zeta^1 - \zeta^2$ and $\zeta^+ \equiv \zeta^1 +\zeta^2 +\psi.$ The first two variations (\ref{eqn:v1}) and (\ref{eqn:v2}) correspond to two vector multiplets and the final variation (\ref{eqn:h}) to a single hypermultiplet. It should be noted however that this last variation does not include all of the necessary field content to fill out the entire hypermultiplet\footnote{In particular, there should be a total of four scalars in this hypermultiplet. Evidently in the truncation these scalars are already fixed to constant values.}.  Due to the appearance of a bare gauge field $(c_1)$ in the hypermultiplet variation it is evident that, in the $AdS$ vacuum, the vector multiplet containing $c_1$ has ``eaten" this hyper. Away from the $AdS$ critical point, one can introduce a Stuckelburg scalar which is charged under $c_1$ and acts as one of the missing hyperscalars. It would be interesting to pursue a more generic reduction which includes all the necessary field content to be supersymmetric off-shell, we leave this for future work.

\subsection{$AdS_5$ solutions}\label{appsec:ads}

$AdS_5$ solutions to this system have been studied for $z=0$ and $z=1$ in \cite{Maldacena:2000mw,Gaiotto:2009gz,Benini:2009mz}. Recently an infinite class of solutions were found in \cite{Bah:2011vv,Bah:2012dg}. For $\kappa = -1$ these are valid solutions of the equations of motion for arbitrary values of $z.$ For $\kappa = 1$ and $\kappa = 0$ the solutions are only valid for $|z| > 1$ and $z \neq 0,$ respectively.\footnote{This can be seen directly in the last equation in (\ref{eq:Pext}) which is a sum of non-vanishing positive terms if $z$ does not satisfy the condition $|z| > 1$ for $\kappa = 1$ and $z\neq0$ for $\kappa =0.$}

In the $AdS_5$ vacuum we fix all the vectors to zero. The solutions found in \cite{Bah:2011vv,Bah:2012dg} have the five-dimensional metric given by $ds_{AdS_5}^2$ at which point the scalars attain the values\footnote{Note that in these solutions the seven-dimensional supergravity coupling has been fixed as $m=2$. Also recall that in comparing with \cite{Bah:2011vv,Bah:2012dg}for $\kappa = \pm 1,$ $z_{here} = -\kappa z_{there}$}
\begin{eqnarray} \label{eq:AdSsol}
e^{10\lambda_1} &=& \frac{-\kappa(1 - 7\kappa z + 7z^2) +33z^3 - (|\kappa|-4\kappa z+19z^2)\sqrt{|\kappa|+3z^2}}{4z(\kappa+z)^2}\,, \nn \\
e^{2\lambda_1-2\lambda_2} &=& \frac{-\kappa+z}{2z+\sqrt{|\kappa|+3z^2}}\,, \nn \\
e^{2B} &=& -\frac{1}{8}e^{2\lambda_1+2\lambda_2}\left((\kappa+z)e^{2\lambda_1} + (\kappa-z)e^{2\lambda_2} \right)\,.
\end{eqnarray}

\section{Spinor and Gamma Matrix Conventions}\label{sec:gammaconv}

Here we list some details of our spinor and gamma matrix conventions. The seven-dimensional gamma matrices $\tilde\gamma_M$ are decomposed as
\begin{equation}
\tilde\gamma_\mu = \gamma_\mu \otimes \sigma_3\,, \qquad \tilde\gamma_6 = 1 \otimes \sigma_1\,, \qquad \tilde\gamma_7 = 1 \otimes \sigma_2.
\end{equation}

With these conventions, we can choose the spinor parameter so that $\tilde\gamma^{67}\epsilon = -i\epsilon$ so that $\eta = \btop{0}{1}$ in (\ref{eq:fermred}). In the text we also choose $SO(5)_c$ conventions in which the $SO(5)_c$ gamma matrices take the following eigenvalues on the spinor
\begin{equation}\label{eq:proj}
\Gamma^{12}\epsilon = \Gamma^{34}\epsilon = -i\epsilon
\end{equation}
which imply $\Gamma^5\epsilon = -\epsilon.$

\end{document}